%% file: main.tex
\documentclass[12pt]{article}
\usepackage{amsmath}
\usepackage{amssymb}
\usepackage{amsthm}
\usepackage{graphicx,psfrag,epsf}
\usepackage{enumerate}
\usepackage{natbib}
\usepackage{algorithm2e}
\usepackage{xcolor}
\usepackage{setspace}
\usepackage{mathtools}

\usepackage{float} 
\usepackage{pifont}
\usepackage{hyperref}
\hypersetup{
    colorlinks=true,
    linkcolor=blue,
    filecolor=blue,      
    urlcolor=blue,
    citecolor=blue
}

\usepackage{authblk}            

\usepackage{xparse}   

\usepackage{booktabs}

\definecolor{burntorange}{HTML}{BF5700}

\definecolor{darkcerulean}{rgb}{0.03, 0.27, 0.49}
\definecolor{darkcandyapplered}{rgb}{0.64, 0.0, 0.0}

\NewDocumentCommand{\rot}{O{45} O{1em} m}{\makebox[#2][l]{\rotatebox{#1}{#3}}}%

\newcommand{\cmark}{\textcolor{darkcerulean}{\ding{51}}}%
\newcommand{\xmark}{\textcolor{darkcandyapplered}{\ding{55}}}%

\usepackage{url} 

\newcommand{\blind}{0}

\newcommand{\I}[0]{\mathcal{I}}
\newcommand{\N}[0]{\mathrm{N}}
\newcommand{\trans}{\intercal}

\newcommand{\Xt}{X_{\phi}}
\newcommand{\Wt}{W_{\phi}}

\newcommand{\taut}{\tau_\phi}

\newcommand{\E}{\text{E}}
\newcommand{\cov}{\text{cov}}
\newcommand{\var}{\text{var}}

\newcommand{\betat}{\beta_{\phi}}
\newcommand{\psit}{\psi_{\phi}}

\newcommand{\ezx}{e_{ZX}}
\newcommand{\ezxt}{e_{Z\Xt}}

\DeclareMathOperator*{\argmin}{arg\,min}

\begin{document}

\def\spacingset#1{\renewcommand{\baselinestretch}%
{#1}\small\normalsize} \spacingset{1}



\newcommand{\mytitle}{
Bayesian inference for treatment effects under nested subsets of controls
}

\if0\blind
{
  \title{\bf \mytitle}
   
  

  \author[1]{Spencer Woody}

  \author[2,3]{Carlos Carvalho}

  \author[2,3]{Jared Murray\thanks{Corresponding author. Email to
  \texttt{jared.murray@mccombs.utexas.edu}}}

  \affil[1]{Department of Integrative Biology, University~of~Texas~at~Austin}

    \affil[2]{Department of Information, Risk, and Operations Management, University~of~Texas~at~Austin}

  \affil[3]{Department of Statistics and Data Sciences, University~of~Texas~at~Austin}

  \maketitle
} \fi

\if1\blind
{
  \bigskip
  \bigskip
  \bigskip
  \begin{center}
    {\LARGE\bf \mytitle}


    \bigskip
    \today
    
\end{center}
  \medskip
} \fi


\begin{abstract}
  When constructing a model to estimate the causal effect of a
  treatment, it is necessary to control for other factors which may
  have confounding effects.  Because the ignorability assumption is
  not testable, however, it is usually unclear which minimal set of controls
  is appropriate -- as is their appropriate functional form in the model -- and effect estimation can be sensitive to these
  choices.  A common approach in this case is to fit several models,
  each with a different control specification (under the assumption that the available controls are sufficient but possibly not all necessary to deconfound the treatment effect), but it is difficult to
  reconcile inference for the treatment effect under the multiple resulting posterior
  distributions.  Therefore we propose a
  two-stage approach to measure the sensitivity of effect estimation
  with respect to control specification.  In the first stage, a model
  is fit with all available controls using a prior carefully selected
  to adjust for confounding.  In the second stage, posterior
  distributions are calculated for the treatment effect under
  submodels of nested sets of controls using projected posteriors under the full model, providing valid Bayesian inference.  We demonstrate how our approach
  can be used to detect influential confounders in a dataset,
  and apply it in a sensitivity analysis of an observational study
  measuring the effect of legalized abortion on crime rates.
\end{abstract}

\noindent%
\vfill


\newpage
\spacingset{1.8} 

\input{input/introduction}

\input{input/projected-posteriors}

\input{input/application}
\input{input/path}

\input{input/simulations}

\input{input/discussion}

\appendix



\singlespacing
\bibliography{main}
\doublespacing

\end{document}

%% file: input/introduction.tex
\section{Introduction}

This paper considers estimation of the average or homogenous causal effect of a treatment variable $Z$ on a continuous outcome $Y$ from observational data using regression adjustment.    This amounts to encoding a set of control variables in a design matrix $X$ in the outcome model, so that the full model for the vector of outcomes becomes
\begin{align}\label{eq:linear-model} Y = \tau Z + X\beta + \epsilon, \quad \epsilon \sim \N(0, \sigma^2_\epsilon \mathcal{I})
\end{align}
Throughout the paper we will loosely refer to the columns of $X$ as ``controls'' or ``control variables'', recognizing that these may be transformed, interacted, or non-linear versions of the original variables.  While we focus on homogensous (or average) effects here, the methods we develop apply as well to estimating heterogeneous effects and we discuss this exension in Section~\ref{sec:discussion}.

We will also assume that necessary conditions for interpreting $\tau$ as a causal effect are met, such as overlap, consistency/stable unit treatment values (SUTVA), and strong ignorability (see e.g. \cite{imbens_rubin_2015}). Chief among these is the assumption that a sufficient set of variables have been have been measured to eliminate the bias due to confounders. Confounders are pre-treatment variables that causally influence both the treatment and the outcome.  Exclusion of confounders in the regression model above would generally result in biased estimation of the treatment effect. Throughout this paper we assume that all confounders are measured. 

However, even under this assumption there often remains uncertainty about {\em which} measured variables are confounders -- i.e., whether a susbet of the variables are sufficient to deconfound the treatment effect -- and the most appropriate way to include them in the outcome model (as nonlinear terms, interacted with other variables, and so on). It has long been recognized that these choices can be consequential for estimating $\tau$, and papers using regression adjustment commonly report estimates for the partial effect of $Z$ under varying specifications for $X$ as informal robustness checks. However, these results are difficult to interpret: The typical standard errors, confidence intervals, and p-values (and their Bayesian analogues) in this collection of regression tables are all computed assuming a fixed specification for $X$.

In this paper we provide a Baysesian framework for addressing these questions via posterior summarization. In doing so we obtain proper Bayesian inference simultaneously over effect estimates under different sets of controls, allowing us to conduct a principled sensitivity analysis to different control specifications and formalizing ad-hoc procedures like refitting models and comparing estimates and uncertainty intervals. We also propose a method for exploring which confounders appear the most (or least) ``important''; while insufficient for formal confounder selection (which would require some knowledge of causal relationships {\em among} the controls) our method provides  insight into which variables, higher order terms, and interactions may be more or less important for deconfounding effect estimates.

The paper proceeds as follows: In the remainder of this section we collect background material and related work. In Section~\ref{sec:proj-post-treatm} we introduce our method for obtaining valid Bayesian inference for the treatment effect under nested subsets of the original controls.
In Section~\ref{sec:application} we present an empirical study using observational data from \cite{Levitt} investigating the hypothesis that legalized abortion reduces homicide rates.  Though the original paper provides support in favor of this hypothesis, several subsequent reanalyses have claimed to negate their finding after considering an expanded set of controls.  Using our methods we conduct a sensitivity analysis to check the robustness of \citeauthor{Levitt}'s original finding with respect to these control specifications. In Section~\ref{sec:path-betw-subm} we introduce a semi-automated method to rank control terms according to their apparent (lack of) influence on the effect estimate and apply this method to the Donahue and Levitt data. Section~\ref{sec:discussion} concludes with a discussion and future work.

\subsection{Handling control specification uncertainty under identification}

As noted above, even when we assume all the confounding variables are measured there remains uncertainty about {\em which} measured variables are confounders as well as the appropriate way to include them in the outcome model. Broadly speaking, there are two distinct approaches to this problem.  One approach, especially common in the econometrics literature, is to fit several models  with different sets of controls and inspect how estimates of the treatment effect vary under each one.  This is effectively an ad hoc sensitivity analysis investigating whether the treatment effect is robust to specification of $X$.  {For instance, \cite{Levitt}, whose application we revisit in Section~\ref{sec:application}, estimate the effect of abortion on crime rates in the United States by fitting multiple models, at times adding or removing certain control terms such as state-level fixed effects.  They then inspect how concordant these estimates are between the models.  }

\cite{speccurve} introduce some formalism to this approach with the \textit{specification curve}, whereby a large multitude of ``reasonable'' control specifications are pre-specified and inference for the treatment effect is considered across each of these.  \cite{speccurve} also provide a valid joint test for null treatment effects across all specifications. This test inherits the drawbacks of null hypothesis significance testing generally and more importantly is rather limited (for example, it fails to address estimation questions, e.g. how the effect estimate changes from one specification to another).  Further, the universe of regressions specifications is itself a researcher degree of freedom. Using expansive sets of controls with nonlinear and interactive terms (without justification) can introduce sufficient sampling variability to ``wash away'' treatment effects \citep*{bryan2019}. 

Neither the ad-hoc nor specification curve approaches completely address the difficulty in reconciling inference under alternative control specifications, essentially because they are incomplete treatments of model uncertainty.
%
Another approach to the problem is to cast uncertainty about the control specification as a model selection problem, prioritizing the selection of controls which appear to be determinant of both the treatment and the outcome.  For instance, \cite{belloni2014inference} introduces a method giving valid inference on the treatment effect after selection of a sparse set of confounders are found using the LASSO.  \cite{wilson2014confounder} use penalized high-posterior density intervals to select confounders. 

Some Bayesian models utilize selection/exposure models along with outcome models to this end:
\begin{align}\label{eq:exp}
  \begin{split} \text{Exposure model:} \quad (Z \mid X) &= X\gamma + \nu, \quad \nu \sim \mathcal{N}(0, \sigma_\nu^2)\\ \text{Outcome model:} \quad (Y \mid Z, X) &= \tau Z + X\beta + \epsilon, \quad \epsilon \sim \mathcal{N}(0, \sigma_\epsilon^2).
  \end{split}
\end{align} 
One can attempt to apply shrinkage or selection that prioritizes variables predictive of both exposures and outcomes.  \cite*{wang2012bayesian} introduce a variant of Bayesian model averaging to estimate $\tau$ in this way to account for uncertainty in the set of confounders.  They tune their spike-and-slab prior to favor inclusion of variables in the outcome model if they appear in the exposure model, i.e., $\beta_j$ is more likely to be non-zero if $\gamma_j$ is nonzero.  
This approach was extended to generalized linear models for describing outcome and exposure, and to allow for treatment effect heterogeneity, by \cite*{wang2015}. 

\cite*{RIC} use a similar parameterization after applying the transformation $(\tau, \beta + \tau \gamma, \gamma) \rightarrow (\tau, \beta_d, \beta_c)$ to Eq.~\eqref{eq:exp}, thus giving
\begin{align}\label{eq:ric}
  \begin{split} \text{Exposure model:} \quad (Z \mid X) &= X\beta_c + \nu, \quad \nu \sim \mathcal{N}(0, \sigma_\nu^2)\\ \text{Outcome model:} \quad (Y \mid Z, X) &= \tau (Z - X\beta_c) + X\beta_d + \epsilon, \quad \epsilon \sim \mathcal{N}(0, \sigma_\epsilon^2),
  \end{split}
\end{align}
and then specify independent shrinkage priors for $\beta_d$ and $\beta_c$.  This approach wards off ``regularization-induced confounding,'' i.e., the bias in estimating $\tau$ resulting from the naive use of shrinkage priors on $\beta$ in Eq.~\eqref{eq:linear-model}.  Their approach also has a computational advantage over that of \cite{wang2012bayesian} in that posterior sampling is often easier when using continuous shrinkage priors than it is for spike-and-slab type priors.  However, even with careful regularization, inclusion of a large number of controls can still drown out any signal in the treatment effect if the number of observations is insufficient.

In this paper we propose what is essentially a hybrid approach between the ad-hoc sensitivity analysis and model selection perspectives, first fitting a large model that includes a broad set of controls that are sufficient to deconfound the treatment effect, utilizing shrinkage or selection as necessary. This first-stage model should adequately represent model uncertainty. Our particular analyses below are based on the shrinkage priors of \cite{hahn2018bayesian}, but the methods we develop are agnostic to the specific choice of model and prior. 

In the second stage we use the fitted posterior distribution to assess sensitivity to control specifications, without re-using the original data to fit a new model. This is a statistically principled version of the ad-hoc sensitivity analysis which is not limited to a predetermined set of specifications and furnishes valid statistical inference over how estimates vary across control specifications. It is related to the notion of posterior summarization \citep{woody2019model}, which we describe below.

\subsection{Posterior Summarization}\label{sec:post-summ-review}

Our methods are most closely related to previous work on posterior summarization, most notably \cite*{woody2019model} who introduce methods for producing interpretable summaries of global and local predictive behavior of Bayesian nonparametric regression models, \cite{DSS}, who create sparse coefficient summaries of high-dimensional linear models, and \cite{maceachern2001decision}, who first introduced linear summaries of nonlinear regression models.  The idea is to approximate a fitted high-dimensional or otherwise complex regression function by a simpler surrogate (the ``summary''). 

Concretely, suppose we have the following model for predicting $Y$  using covariates $W$:
\[
Y = f(W) + \epsilon.
\]
\cite*{woody2019model} propose examining the posterior distribution of  interpretable approximations to $f$ (additive, linear, sparse, and so on) which we denote by $\gamma\in \Gamma$. These approximations are obtained as the minimizers of a loss function
\begin{align}\label{eq:general-loss-fn}
  \mathcal{L}(f, \gamma, W^*) &= d(f, \gamma, W^*) + q_\lambda(\gamma),
\end{align}
where $d(\cdot, \cdot, W^*)$ measures discrepancy between
between the original regression function $f$ and the 
summary  $\gamma$ over the some specified covariate locations of
interest $W^*$. The second term $q_\lambda(\cdot)$ is an optional penalty to discourage complexity in the summary.

By examining the posterior distribution of
\begin{align}\label{eq:general-loss-min}
  \argmin_{\gamma\in  \Gamma} \mathcal{L}(f, \gamma, W^*) &= d(f, \gamma, W^*) + q_\lambda(\gamma),
\end{align}
one obtains the posterior distribution of the ``best'' summary of $f$ (according to \eqref{eq:general-loss-fn}) -- for example, this might be the best linear approximation to $f$ as measured by squared error. Since the data are used one time -- to obtain the full posterior -- one is free to consider as many posterior summaries as desired, varying the class of summaries or any component of the loss funciton in Eq.~\eqref{eq:general-loss-fn}, without compromising the validity of the resulting inferences.

Posterior summarization has been applied in a range of settings, including variable selection summaries in seemingly unrelated regressions \citep{PuelzSUR, Chakraborty2016}, sparse portfolio selection \citep{optimalETF, PuelzPortfolio}, estimation of sparse precision matrices \citep{bashir2018post}, and factor analysis \cite{bolfarine2021decoupling}.  Additionally, this approach has been applied to variable selection in functional regression \citep{kowal2018bayesian} and variable selection under economic considerations \citep{maceachern2019economic}.  \cite{crawford2019predictor} produce linear summaries for nonparametric regressions followed by variable selection, and, similarly, \cite{lee2014inference} calculate posteriors for M-estimators as summaries for nonparametric density estimates. This paper diverges from earlier work by emphasizing estimation of treatment effects (or, more generally, partial effects) rather than predictive models or joint distributions/covariance structures. In this paper our summaries will be given by subsets of the control terms, in which case the summary will provide inference on the partial effect of $Z$ under a different set of controls.

%% file: input/projected-posteriors.tex
\section{Posterior summaries and the projected posterior for treatment effects}
\label{sec:proj-post-treatm}

We begin by defining a new estimand for assessing sensitivity to the control function with a single model fit: projected regression coefficients. We refer to the posterior of these quantities as the projected posterior. Let $W = [Z \quad X]$ be the concatenated matrix of the treatment vector and controls with corresponding coefficient vector $\psi = [\tau \quad \beta^\trans]^\trans$, so that the outcome model from Eq.~\eqref{eq:exp} becomes
\begin{align}\label{eq:cat}
  Y &= W\psi + \epsilon, \quad \epsilon \sim \mathcal{N}(0,
      \sigma_\epsilon^2 \I). 
\end{align}
Consider a given subset of the controls described by the inclusion vector $\phi \in \{ 0, 1 \}^p$.  This has the corresponding restricted design matrix $\Xt$, the $\phi$-subset of columns of $X$, along with the restricted coefficient vector $\betat$.  Finally, similar to $W$, define $\Wt = [Z \quad \Xt]$. The projected posterior distribution corresponding to this set of controls is the posterior distribution of 
\begin{align}\label{eq:projection}
  \psit \equiv [\taut\quad \betat^\trans]^\trans &= {(\Wt^\trans \Wt)^{-1} \Wt^\trans W} \psi. 
\end{align}
Note that we will occasionally refer to $\taut$ as a causal effect estimate even when the subset of controls given by $\phi$ is not sufficiently rich to eliminate confounding. In this case $\taut$ is still a treatment effect {\em estimate} -- just a biased and inconsistent one.

As an estimand Eq.~\eqref{eq:projection} can be motivated in two ways: First via the framework of posterior summarization developed in \cite{woody2019model}, and second via the analysis of hypothetical future data. It is easy to see that the projected posterior is a posterior summary as described in Section~\ref{sec:post-summ-review}:
In the special case that $f(W) = W\psi$, $\gamma(W) = \Wt\gamma$, $q_\lambda(\phi) = 0$ (no complexity penalty), $W^* = W$ and
\[
d(f, \gamma, W) = \| W\psi - \Wt\gamma \|_{2}^2
\]
it is easy to verify that 
\begin{align}
  \psit &= \arg \min_{{\gamma} \in \mathbb{R}^{|\phi|}}
                    \| W\psi - \Wt\gamma \|_{2}^2.
\end{align}
So $\psit$ is the vector of coefficients in the restricted linear model with the closest predictions to the full model (in squared-error) over the originally observed covariate locations. Equivalently, it is the vector of coefficients in the OLS regression of the ``denoised data'' $\hat{y} = W\psi$ on $\Wt$.

\newcommand{\Yrep}{Y^*}
\newcommand{\Wrep}{W^*}
\newcommand{\Xrep}{X^*}
\newcommand{\Zrep}{Z^*}
\newcommand{\Xtrep}{\Xt^*}
\newcommand{\Wtrep}{{\Wt^{*}}}

We can also derive $\psit$ from a hypothetical repeated analysis of future data. Observe that under the model in Eq.~\eqref{eq:cat} the distribution of future outcomes $\Yrep$ observed at control and treatment values $\Xrep$ and $\Zrep$ (concatenated in $\Wrep$) is given by
\begin{align}
  \Yrep\sim  \mathcal{N}(\Wrep\psi,
      \sigma_\epsilon^2 \I). 
\end{align}
For the subset of controls given by $\phi$ collected in $\Wtrep$, the posterior distribution of the coefficients in the ordinary least squares (OLS) regression of $\Yrep$ on $\Wtrep$ in a future dataset is 
\begin{align}
  \hat{\tilde{\psi}}_{rep} \equiv (\Wtrep^{\trans} \Wtrep)^{-1}\Wtrep^{\trans} \Yrep\sim  \mathcal{N}( (\Wtrep^{\trans} \Wtrep)^{-1}\Wtrep^{\trans} \Wrep\psi,
      \sigma_\epsilon^2 (\Wtrep^{\trans} \Wtrep)^{-1}),
\end{align} 
so when $\Wrep = W$ and $\Wtrep = \Wt$,
\begin{align}\label{eq:postpred}
  \hat{\tilde{\psi}}_{rep} \sim  \mathcal{N}( (\Wt^{\trans} \Wt)^{-1}\Wt^{\trans} W\psi,
      \sigma_\epsilon^2 (\Wt^{\trans} \Wt)^{-1}). 
\end{align}

Therefore $\psit = (\Wt^\trans \Wt)^{-1}\Wt^\trans W\psi$ is the {\em expected value} of the OLS coefficients in a new dataset with identical control values and treatment assignments to those originally observed, given the model parameters $\psi$. (The expectation is with respect to the future outcomes $\Yrep$, the distribution of which is given by the model in Eq.~\ref{eq:cat}.) Note that while we assume independent homoskedastic normal error terms in this paper, due to the linearity of the projection operator this interpretation of the projected posterior holds in general provided the full model has additive mean-zero errors (and light enough tails that the expected value exists).

It is important to distinguish between $\hat{\tilde{\psi}}_{rep}$, the vector of OLS coefficients in a future dataset, and its expected value $\psit$. The former includes irrelevant sampling variability, and often corresponds to an analysis we would not perform if we were actually given this hypothetical future dataset. However, the {\em expected value} of the OLS regression coefficients across replicated datasets is obviously an estimand of interest, since the OLS coefficients are unbiased or at least consistent for interesting (super-)population parameters under mild conditions.\textbf{}

\subsection{Inference using the projected posterior}

The projected posterior lets us make inference about the partial effect of $Z$ under different sets of controls while only using the data once -- to move from prior to posterior in the full model. Clearly then uncertainty statements about $\psit$ and $\taut$ using the projected posterior flow entirely from the posterior for $\psi$ and the other parameters in the full model.  That is, the projected posterior does {\em not} represent uncertainty {\em about} the ``true'' model, but it allows us to make inference about how we expect regression coefficients to change under different control specifications -- provided that we are willing to make inference using the original fitted model. 

Therefore, for the projected posterior to be useful the original fitted model should 1) adequately capture model uncertainty and 2) be judged acceptable for making inference. Regarding 1), in this paper we take that to mean including the union of the ``interesting'' controls in the outcome model and applying shrinkage or selection as necessary to fit the model (taking care to appropriately address confounding \citep{hahn2018bayesian}). The methods proposed here are model- and prior-agnostic, however, and we discuss generalizations in Section~\ref{sec:discussion}.

Provided these two conditions are met, projected posteriors enable inference on a rich set of estimands.  Note that the posterior under the full model implies a full joint posterior distribution for the original model parameters and any {\em set} of projections. This allows us to directly examine quantities like the posterior for $\tau - \taut$, the expected ``bias'' in the partial effect of $z$ resulting from adjusting for $\Xt$ instead of $X$, or the difference in $\taut$ under distinct and possibly non-nested subsets of controls (when both are nested within the full model). This gives us a meaningful way to reason about how we expect regression coefficients to move as we add or drop controls -- with valid measures of uncertainty -- compared to the common practice of refitting models and examining the change in estimated coefficients and their standard errors. 

\subsection{Properties of the projected posterior}\label{sec:properties}

\newcommand{\alphat}{\tilde{\alpha}}
\newcommand{\betac}{\beta_{\neg\phi}}
\newcommand{\Xc}{X_{\neg\phi}}
\newcommand{\zbetac}{\gamma_{\neg\phi}}
\newcommand{\hatcbetat}{\hat{\eta}}
\newcommand{\hatzbetac}{\hat{\gamma}_{\neg\phi}}

We are primarily interested in $\taut$, the regression coefficient on $Z$ using the reduced set of controls. In what follows we will assume that $X$ is full rank with $n>(p+1)$, although many results hold provided only that $\phi$ and its complement contain fewer then $n-1$ controls.

Let $P_{V} = V(V^\trans V)^{-1}V^t$ be the projection matrix onto the column space of $V$ and {${e_{UV} = (I-P_V)U}$} be the vector of residuals in the OLS regression of $U$ on $V$. It is straightforward to show that
\begin{align}
\taut &= \tau + \frac{\ezxt^\trans (I-P_{\Xt})\Xc}{\ezxt^\trans \ezxt}\betac \label{eq:alphat}\\
&= \tau + \frac{{\hatzbetac}^\trans  \Xc^\trans(I-P_{\Xt})\Xc \betac }{\ezxt^\trans \ezxt}\label{eq:alphatpartial}
\end{align}
where $\hatzbetac$ are the coefficients on $\Xc$ in the least squares regression of $Z$ on all of $X$. 
We see immediately that $\taut\approx\tau$ when $\betac$ or $\hatzbetac$ are near zero -- that is, either the estimated partial effects of $\Xc$  on $Y$ are small in our model or the partial effects of $\Xc$ on $Z$ are small in an OLS regression (adjusting for $\Xt$ in both cases). 

Provided that the controls in $\Xt$ are sufficient to deconfound the treatment effect, our model is otherwise well-specified, and the sample size is sufficiently large,  we would expect the former to hold for instrumental variables or terms that ``act'' like instruments. (For example, suppose $\Xt$ contains main effects for two variables and $\Xc$ their interaction which has a nonzero coefficient in the treatment model but not in the outcome model. In this case the interaction term ``acts'' as an instrument, even though the two original variables are not themselves instruments.) Similarly, we would expect the latter to hold for purely prognostic variables or controls. In neither case is the converse true, however -- the mere fact that $\taut\approx\tau$ on removing a control is insufficient evidence to conclude anything about that control's causal relationships with $Z$ or $Y$.

Equations~\eqref{eq:alphat} and $\eqref{eq:alphatpartial}$ tell us immediately how the expected value of $\taut$ relates to the expected value of $\tau$ and $\betac$. Other features of the posterior distribution of $\taut$ depends on the prior and data in a complicated way. Take the simplest case when $\Xc$ contains a single column. We can rewrite Eq.~\eqref{eq:alphat} as the familiar omitted variables formula:
\begin{equation}
\taut = \tau + \hatcbetat\betac,\label{eq:alphaomitted}\\ 
\end{equation}
where $\hatcbetat$ is the coefficient of $Z$ in the OLS regression of $\Xc$ on $Z$ and $\Xt$. The marginal posterior variance of $\taut$ is then
\[
\var(\taut \mid Y) = \var(\tau\mid Y) + \hatcbetat^2\var(\betac\mid Y) + 2\hatcbetat \cov(\tau, \betac\mid Y).
\]
Reasoning about each piece of this formula independently would be misleading. Note that $\hatcbetat$ measures the partial association between $Z$ and $\Xc$ adjusting for $\Xt$, which will also partly determine the posterior variance-covariance of their coefficients in the outcome model under most priors. Further, for priors that induce shrinkage or selection the marginal posterior variance is often not even a reasonable measure of spread -- posteriors can be skewed, multimodal, or contain singularities. 

However, we can derive more properties of the projected posterior in the special case of a flat prior on the coefficients in the full model\textbf{}. This illustrates some of the behavior of the projected posterior and describes how we might expect it to behave under weaker priors or in large samples where the likelihood dominates.

\subsection{The projected posterior under a flat prior}\label{sec:analytical}

Suppose for the moment we use flat priors $\pi(\beta, \tau) \propto 1$, an independent prior on $\sigma^2_\epsilon$, and the outcome model in Eq.~\eqref{eq:linear-model}. In this case the posterior for $\psi = [\tau \quad \beta^\trans]^\trans$ (given $\sigma^2_\epsilon$) is multivariate Gaussian:
\begin{align}
  (\psi \mid Y) & \sim \mathcal{N}(\hat \psi, \sigma_\epsilon^2 (W^\trans W)^{-1}), \nonumber
\end{align}
with $\hat \psi = (W^\trans W)^{-1} W^\trans Y$, the OLS estimate of $\psi$. 
The marginal posterior of $\tau$ is also normal:
\begin{align}
  (\tau \mid Y) & \sim \mathcal{N}\left(\hat \tau, \frac{\sigma_\epsilon^2}{e_{ZX}^\trans e_{ZX}}\right), \nonumber
\end{align}
where $\hat\tau$ is the OLS estimate of $\tau$. 

For a given subset of controls, the projected posterior for the reduced coefficient vector is also multivariate Gaussian with mean
\begin{align}
  \E(\psit \mid Y)
  &= (\Wt^\trans \Wt)^{-1} \Wt^\trans
    {W (W^\trans W)^{-1} W^\trans} Y \nonumber \\
  &= (\Wt^\trans \Wt)^{-1} \Wt^\trans
    P_W Y = (\Wt^\trans \Wt)^{-1} \Wt^\trans Y = \hat{\psit}, \label{eq:proj-mean}
\end{align}
the OLS estimate of $\psit$, and covariance matrix
\begin{align}
  \cov(\psit \mid Y)
  &= \sigma_\epsilon^2 (\Wt^\trans \Wt)^{-1} \Wt^\trans W (W^\trans
    W)^{-1} [(\Wt^\trans \Wt)^{-1} \Wt^\trans W]^\trans \nonumber \\
  &= \sigma_\epsilon^2 (\Wt^\trans \Wt)^{-1} \Wt^\trans P_W \Wt
    (\Wt^\trans \Wt)^{-1} 
    %
    =  \sigma_\epsilon^2 (\Wt^\trans \Wt)^{-1}
    , \label{eq:proj-cov}
\end{align}  
since $\Wt$ is a subset of the columns of $W$. The marginal posterior of $\taut$ is 
\begin{align}
  (\taut \mid Y) & \sim \mathcal{N}\left(\hat \taut, \frac{\sigma_\epsilon^2}{\ezxt^\trans \ezxt} \right), \nonumber
\end{align}
where $\hat{\taut}$ is the OLS estimate of $\taut$ in the regression $Y = \Wt\psit + \epsilon$.

Since $\E(\tau\mid Y) = \E(\E(\tau\mid \sigma^2_\epsilon, Y)) = \hat{\tau}$ and $\E(\taut\mid Y) = \E(\E(\taut\mid \sigma^2_\epsilon, Y)) = \hat{\taut}$ we can rely on standard least-squares regression results to reason about how the location of the original and projected posteriors relate to one another. For example, on removing a purely prognostic or instrumental control we can expect $\E(\tau\mid Y) \approx E(\taut\mid Y)$. In other cases the difference $\hat{\tau} - \hat{\taut}$ is driven by standard results about omitted variables, with exact expressions given by substituting the OLS estimate for $\betac$ under the full set of controls in Equations~\eqref{eq:alphat}, \eqref{eq:alphatpartial} and \eqref{eq:alphaomitted}.

Turning to the posterior variance, we have
\begin{align}
    \var(\tau\mid Y) &= \var( \E(\tau\mid\sigma_\epsilon, Y)\mid Y) + \E(\var(\tau\mid\sigma_\epsilon, Y)\mid Y)\nonumber\\
    &= \var(\hat{\tau}\mid Y) + \E\left(\frac{\sigma_\epsilon^2}{e_{ZX}^\trans e_{ZX}}\mid Y\right) \\ &= \frac{\E(\sigma_\epsilon^2\mid Y)}{e_{ZX}^\trans e_{ZX}}.
\end{align}
A similar calculation gives 
\[
\var(\taut\mid Y) = \frac{\E(\sigma_\epsilon^2\mid Y)}{\ezxt^\trans \ezxt},
\]
so we have
\begin{equation}
    \frac{\var(\tau\mid Y)}{\var(\taut\mid Y)} = \frac{{\ezxt^\trans \ezxt}}{{\ezx^\trans \ezx}} > 1\label{eq:projvarratio}
\end{equation}
for any independent prior on $\sigma^2_\epsilon$. The posterior variance of $\taut$ is no larger than the variance of $\tau$, and the stronger the partial association between $Z$ and $\Xc$ adjusting for $\Xt$ the greater the reduction in variance. 

At first blush this reduction in variance might seem surprising, but it is actually quite natural: If $\Xc$ is predictive of $Z$ after adjusting for $\Xt$, dropping it reduces multicollinearity leading to a consummate reduction in uncertainty about the partial effect of $Z$ in the reduced set of controls. But because we are using the projected posterior (rather than e.g. refitting a model with solely $\Wt$) we are still leveraging the partial association $\Xc$ has with $Y$ to reduce uncertainty represented by $\sigma^2_\epsilon$ in the original model, which propagates to the projected posterior. Essentially we get inference about the regression using only $\Xt$ as controls while still using all of $X$ to ``denoise'' $Y$.

Compare this result to what we would obtain from refitting the model using only $\Wt$:
\begin{align}
  \nonumber
  Y = \Wt \psit + \epsilon', \quad \epsilon' \sim \N(0, \sigma_{\epsilon'}^2\I).
\end{align}
using the same flat priors. The posterior for $\psit$ in this case is also multivariate
Gaussian with mean vector $\hat{\psit}$ 
%
%
and covariance matrix
\begin{align}\nonumber
  \cov_\text{refit}(\psit \mid Y) = \sigma_{\epsilon'}^2(\Wt^\trans \Wt)^{-1},
\end{align}
and the marginal posterior variance for $\taut$ in the refitted model is
\[
\var_{\text{refit}}(\taut\mid Y) = \frac{\E_{\text{refit}}(\sigma_{\epsilon'}^2 \mid Y)}{\ezxt^\trans \ezxt},
\]
so we have
\begin{align}
    \frac{\var(\tau\mid Y)}{\var_\text{refit}(\taut\mid Y)} &= \frac{\E(\sigma_{\epsilon}^2 \mid Y)}{\E_{\text{refit}}(\sigma_{\epsilon'}^2 \mid Y)}\frac{{\ezxt^\trans \ezxt}}{{\ezx^\trans \ezx}}\label{eq:refitvarratio}\\
    \frac{\var(\taut\mid Y)}{\var_\text{refit}(\taut\mid Y)} &= \frac{\E(\sigma_{\epsilon}^2 \mid Y)}{\E_{\text{refit}}(\sigma_{\epsilon'}^2 \mid Y)}.\label{eq:projrefitvarratio}
\end{align}

With large samples and a prior that leads to consistent estimates of the error variance, we would expect $\E_{\text{refit}}(\sigma_{\epsilon'}^2 \mid Y)\geq \E(\sigma_{\epsilon}^2 \mid Y)$, with near equality when $\Xc$ comprises ``noise'' or instrumental controls (which have no partial effects on $Y$ adjusting for $\Xt$) as they will not reduce the outcome error variance. These are precisely the controls one might like to remove, as they do not correct any bias due to confounding and contribute only uncertainty about the partial effect of $Z$. We could also remove purely prognostic controls without biasing the treatment effect estimate. However, unlike the projected posterior, in the refitted model we pay a price in posterior variance since we expect $\E_{\text{refit}}(\sigma_{\epsilon'}^2 \mid Y)> \E(\sigma_{\epsilon}^2 \mid Y)$ in this case.


Overall, the refitted model really has nothing to recommend it over the projected posterior: First and foremost, the uncertainty statements derived from the refitted posterior are not Bayesianly valid, as they assume a fixed model and control specification when there are at least two control specifications under consideration. Even if we chose to ignore this issue a refitted model yields less precise estimates than the projected posterior when removing purely prognostic controls, and about the same variance reduction as the projected posterior when removing noise or instrumental controls. Finally, the projected posterior has the added benefit of providing a joint posterior for $\tau$ and the projected treatment effects under {\em any} set of controls, which lets us directly make inference on whether two estimated partial effects of $Z$ differ using different sets of controls.

%% file: input/application.tex
\section{A sensitivity analysis of the impact of abortion on crime}
\label{sec:application}

In a high-profile analysis, \cite{Levitt} provided evidence that legalized abortion reduces the number of unwanted children, who are more likely to become criminals, and thereby lowers crime rates in society.  They consider violent crime, property crime, and homicide, and have measurements of these crime rates in each U.S. state for the years 1985 through 1997.  The treatment variable $Z$ is the ``effective'' abortion rate, weighted by criminal age at the time of the arrest.  For the control variables $X$, the authors include eight state-level variables for each year that could also contribute to crime rates, including
\begin{itemize}
\item log-prisoners per capita
\item log-police per capita
\item state unemployment rate
\item log-state income per capita
\item poverty rate
\item level of monetary assistance from Aid to Families with Dependent Children (AFDC), lagged by fifteen years
\item an indicator for presence of a concealed weapons law, and 
\item beer consumption per capita.
\end{itemize}
They also allow for state- and year-level fixed effects with the inclusion of dummy variables.  After removing observations for Alaska, Hawaii, and the District of Columbia there are $p_1 = 67$ total control variables with $n = 624$ observations.  The authors use linear regression to claim that abortion has a significant negative effect on all three of these crime rates.

The provocative nature of this result has attracted scrutiny of their methods.  In particular, several reanalyses \citep{FooteGoetz,belloni2014inference,RIC}, which expand the model to allow for interactions and nonlinear functional forms for the controls, claim to refute the original finding by \citeauthor{Levitt}. The discrepancy between the original published results and those of the subsequent reanalyses using the same potential confounders but varying the specificaiton of the controls makes this a natural setting for us to conduct a sensitivity analysis.

Here we focus only on studying the impact of abortion on the homicide rate.  First, we estimate the treatment effect under a ``large'' model with an expanded set of controls.  Then, we consider inference on the treatment effect under several increasingly parsimonious subsets of these controls by gradually removing terms, and calculate the projected posterior for the treatment effect under these specified subsets.  At the end we compare the treatment effect estimates across these control subsets to gauge the robustness of the original finding of a negative effect.

The ``large'' model we construct is the same as that from \cite{RIC}, who, in addition to the original state-level control variables and state- and year-level fixed effects in the model originally specified by \cite{Levitt}, also include several interactive terms, namely
\begin{itemize}
\item interactions between the eight state-level controls and
  year
\item interactions between the eight state-level controls and
  year-squared
\item interactions between state dummy variables and year, and
\item interactions between state dummy variables and year-squared.
\end{itemize}

This expands the model to include $p_2 = 176$ terms, and allows for quadratic temporal trends for each state and for the effect of the state-level controls.  Note that this does not introduce new confounding variables, but dramatically changes the functional form and dimension of the control function.  In their paper, \cite{RIC} claim that the causal effect of abortion on crime disappears after using this augmented set of controls.  They use a flat prior for the treatment effect $\tau$ and the horseshoe prior \citep{horseshoe} for the coefficients $\beta_c$ and $\beta_d$ in the model given by Eq.~\eqref{eq:ric}, which reparameterizes the likelihood to prevent what they term ``regularization induced confounding.'' We obtain the posterior for the treatment effect $\tau$ and outcome regression coefficients $\beta$ using the same prior specification from \cite{RIC}, and then use this posterior to calculate the projected posteriors found by gradually dropping the control interaction terms.

The control subsets we consider are presented in Table~\ref{tab:control-spec}.  Figure~\ref{fig:levitt-submodel-proj-comp} presents the projected posteriors for each of these, with the columns indexed to match the rows of the table of control subsets.  For the sake of comparison, for each restricted control set, we also show the posterior for the treatment effect found from refitting using a flat prior, along with the posterior found from refitting using the horseshoe prior with the RIC parameterization developed by \cite{RIC}.

\begin{table}[ht!]
  \centering
  \begin{tabular}{lccccccc}
    Control subset & \rot[45]{State controls}
                   & \rot[45]{State dummies}
                   & \rot[45]{Year dummies}
                   & \rot[45]{State controls $\times$ year}
                   & \rot[45]{State dummies $\times$ year}
                   & \rot[45]{State  controls $\times$ year$^2$}
                   & \rot[45]{State dummies $\times$ year$^2$}                    \\
    \toprule
    0$^*$      & \cmark & \cmark & \cmark & \cmark & \cmark & \cmark & \cmark \\ \midrule 
    1.1            & \cmark & \cmark & \cmark & \cmark & \cmark & \cmark & \xmark \\ \midrule
    1.2            & \cmark & \cmark & \cmark & \cmark & \cmark & \xmark & \cmark \\ \midrule
    2              & \cmark & \cmark & \cmark & \cmark & \cmark & \xmark & \xmark \\ \midrule
    3.1            & \cmark & \cmark & \cmark & \cmark & \xmark & \xmark & \xmark \\ \midrule
    3.2            & \cmark & \cmark & \cmark & \xmark & \cmark & \xmark & \xmark \\ \midrule 
    4$^\dag$       & \cmark & \cmark & \cmark & \xmark & \xmark & \xmark & \xmark \\ \midrule 
    5.1            & \cmark & \cmark & \xmark & \xmark & \xmark & \xmark & \xmark \\ \midrule 
    5.2            & \cmark & \xmark & \cmark & \xmark & \xmark & \xmark & \xmark \\ \midrule
    6              & \cmark & \xmark & \xmark & \xmark & \xmark & \xmark & \xmark \\ \midrule
    7              & \xmark & \cmark & \cmark & \xmark & \xmark & \xmark & \xmark \\ \midrule
    8$^\ddag$      & \xmark & \xmark & \xmark & \xmark & \xmark & \xmark & \xmark \\
    \bottomrule
\end{tabular}
\caption{Control subsets used for calculation of projected posteriors, shown in Figure~\ref{fig:levitt-submodel-proj-comp}.  $*$: Control specification from \cite{RIC}; $\dag$: control specification from \cite{Levitt}; $\ddag$: null set of controls (considering treatment only). }
\label{tab:control-spec}
\end{table}

\subsection{Results of the sensitivity analysis}

\begin{figure}[H]
  \centering
  \includegraphics[height=0.35\textheight]
  {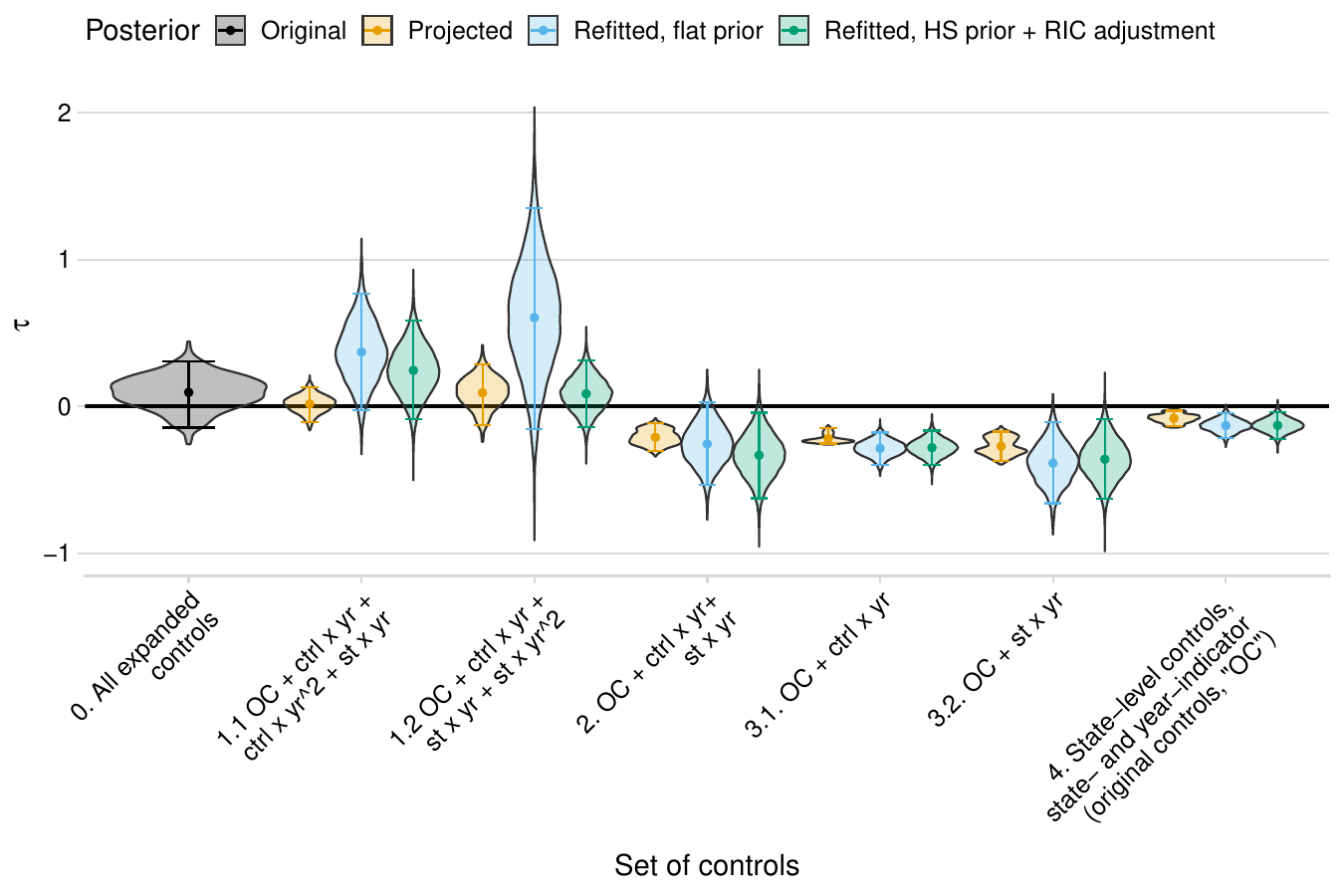}\\
  \includegraphics[height=0.35\textheight]
  {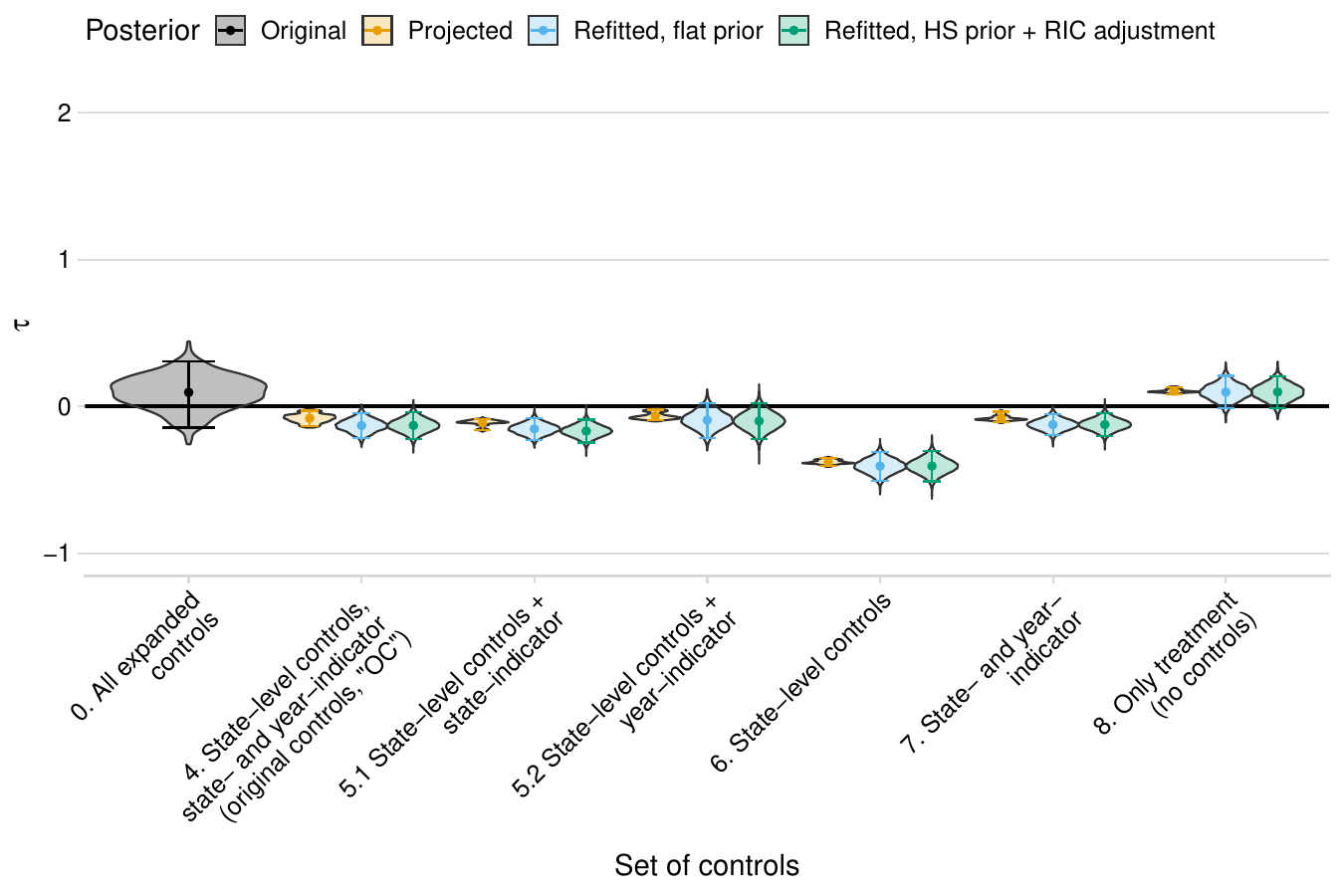}
  \caption{\label{fig:levitt-submodel-proj-comp} Projected posteriors for selected nested summaries, along with posteriors from refitting the models using a flat prior (without any RIC adjustment) and using a horseshoe prior using an RIC adjustment \citep{RIC}, using homicide data from \cite{Levitt}.  Column 4 corresponds to the control specification from the original paper.  }
\end{figure}

\begin{figure}[H]
  \centering
  \includegraphics[width=0.8\textwidth]
  {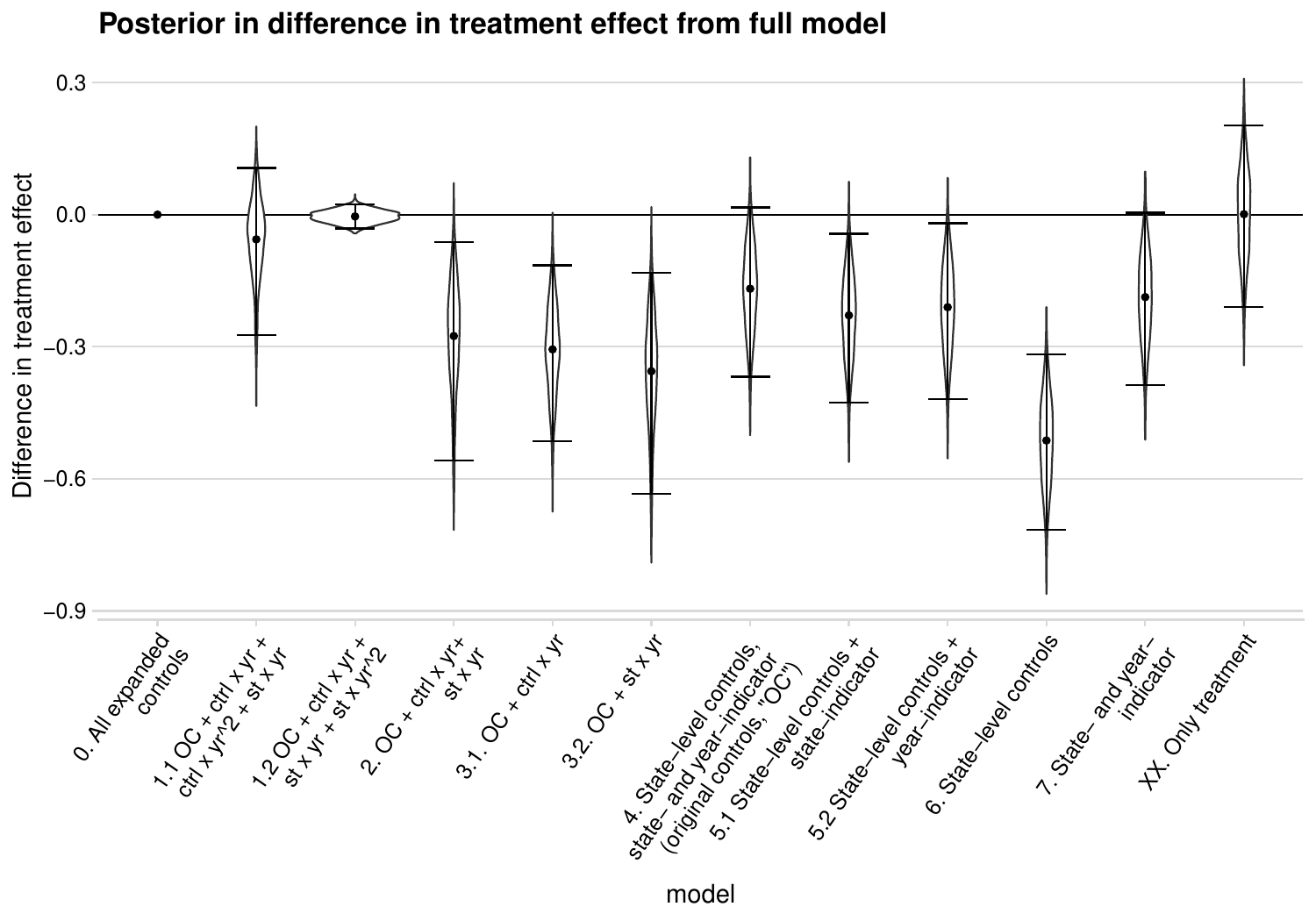}
  \includegraphics[width=0.8\textwidth]
  {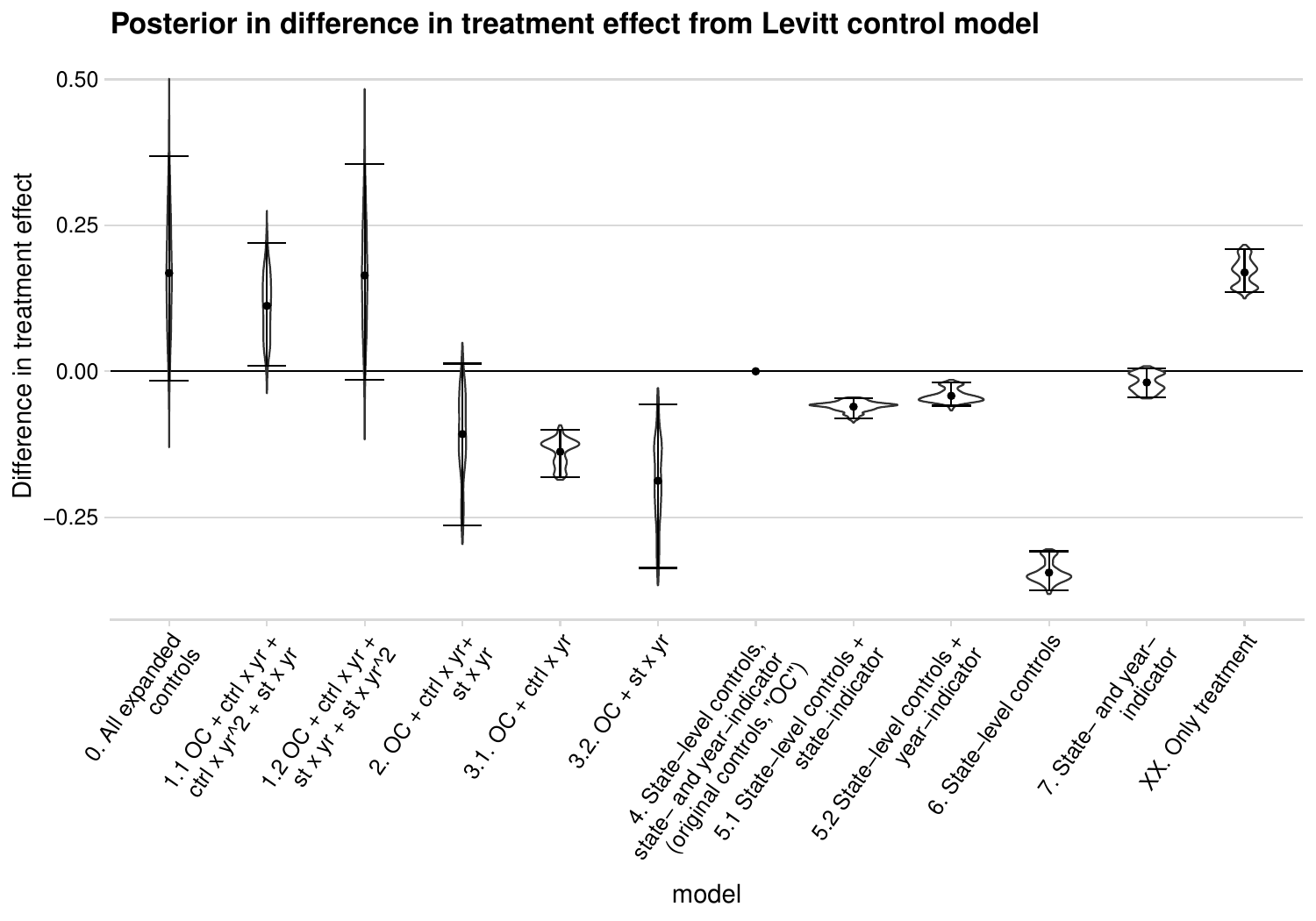}
  \caption{\label{fig:diffplots}
    Posterior for difference in treatment effect under various model specifications and either (i) full model (\textit{top}) or (ii) the model under the original control specification by Donahue and Levitt (\textit{bottom}).
  }
\end{figure}

Our results are collected in Figures~\ref{fig:levitt-submodel-proj-comp} and~\ref{fig:diffplots}. Figure~\ref{fig:levitt-submodel-proj-comp} presents projected posteriors for the effect estimate under each specification, along with the refitted estimates for reference. The top panel shows a path from the full model to the set of original controls, and the bottom panel shows a path from the original set of controls to the naive unadjusted estimate. 
Since the projected and full model effect estimates are dependent in their joint posterior distribution, comparing the intervals in Fig.~\ref{fig:levitt-submodel-proj-comp} would be misleading.
Therefore in Figure~\ref{fig:diffplots} we plot the difference between projected effect estimates under each specification and the estimate from the full model (top panel) or the set of original controls (bottom panel). 

As expected based on the controversy following \cite{Levitt}, the effect estimate is sensitive to the control specification. But our sensitivity analysis sheds new light on how and why this is true. There are several new findings:

\noindent{\bf{The original finding is (largely) directionally robust.}} Only four specifications lead to positive (projected) effect estimates: the full model, the two largest sets of controls (1.1 and 1.2), and the naive unadjusted estimate (8). Examining the posterior for the effect differences between these sets and the original set of controls  (Fig~\ref{fig:diffplots}) we see that for the large specifications there is substantial uncertainty about the size of the effect difference. The remaining specifications lead to effect estimates that are larger in magnitude than the original finding.

\noindent{\bf{The sign of the effect estimate changes upon introducing quadratic trends in time interacted with state dummies or the state-level control variables.}} Specifications 0, 1.1, and 1.2 all share this feature. Given this, it is reasonable to ask whether these quadratic terms can be well-estimated, especially given that 1) the model already includes dummy variables for each year, 2) there are only thirteen years of data available, and 3) many of the controls have time trends themselves and are correlated with the year variable.  We should also ask whether these controls represent substantively defensible assumptions -- for example, do we expect the partial effect of the poverty rate on crime to vary {\em nonlinearly} over these thirteen years? If not, perhaps these control terms should be disregarded. Using the projected posterior, we can remove these terms from our {\em analysis} -- without using our data twice or collecting new data.

\noindent{\bf{The quadratic year by original interaction terms appear to act as instruments in the outcome regression, given all the other controls.}} Figure~\ref{fig:levitt-submodel-proj-comp} shows that the full model and the projected estimate in specification 1.2 have nearly identical point estimates, but the posterior spread of the projected estimate is lower. Figure~\ref{fig:diffplots} confirms that the difference is near zero with a high degree of certainty. Taken together, these suggest that the quadratic year by control variable terms may be acting as instruments, reasoning from our findings in  Section~\ref{sec:analytical}.

\begin{figure}[H]
  \centering
  \includegraphics[width=\textwidth]{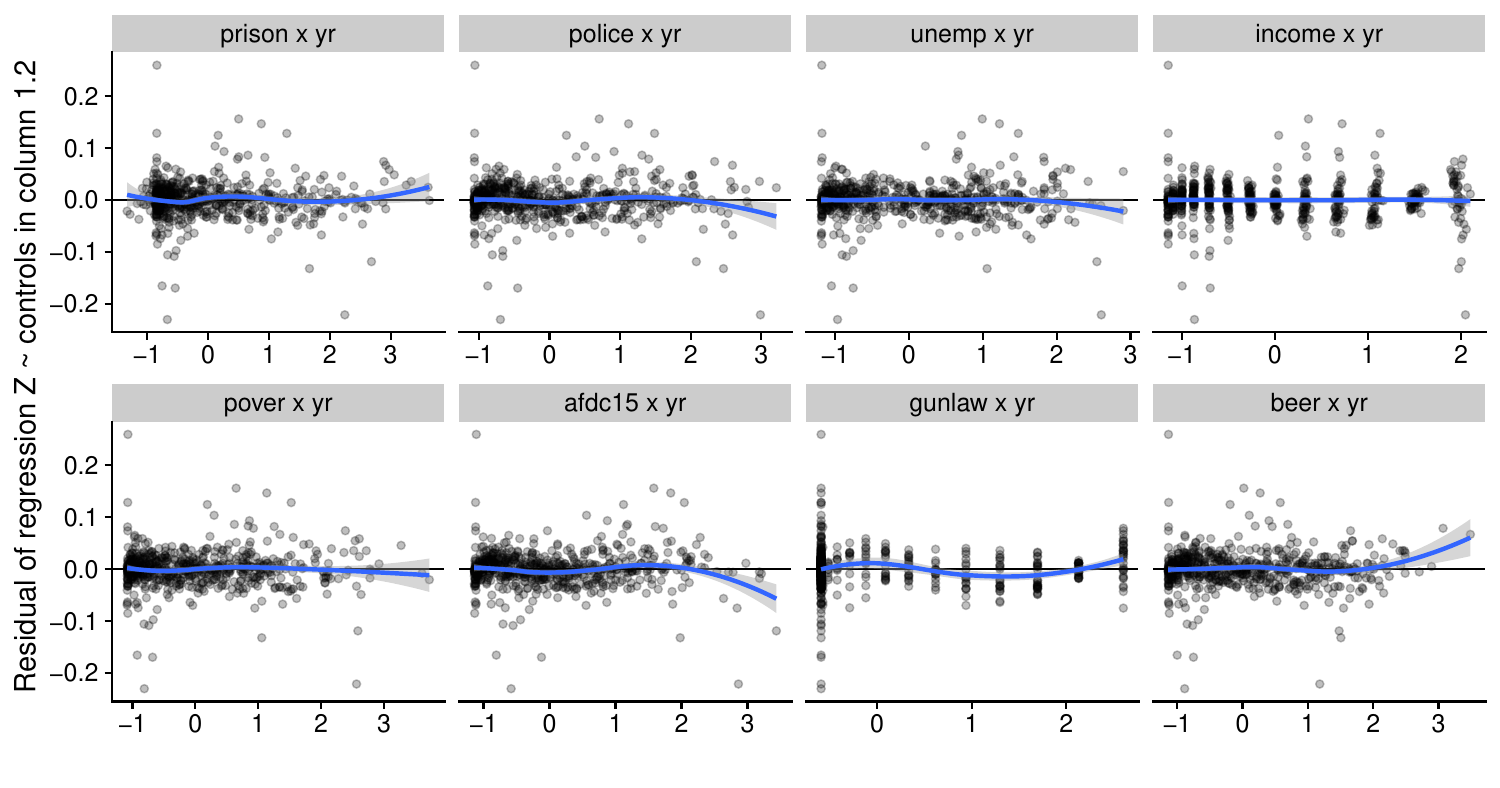}\\
  \includegraphics[width=\textwidth]{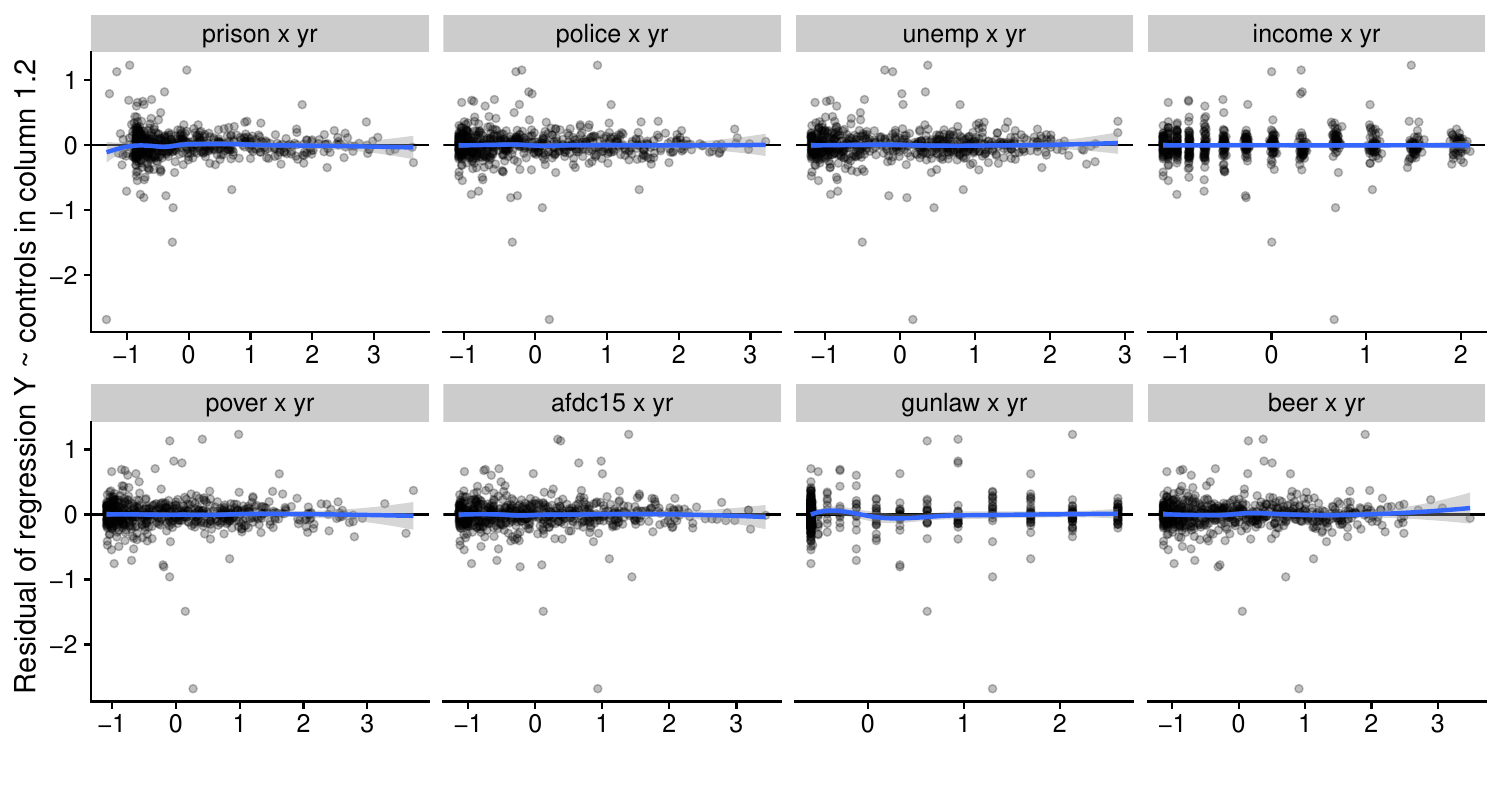}
  \caption{\label{fig:reg-resid}
    Residuals of regression on $Z$ (top) and $Y$ (bottom), with overlaid LOESS smoother curves.  There are more nonlinear trends in the residuals for the $Z$ regression than for the $Y$ regression, suggesting that the control $\times$ year$^2$ variables are instrument-like: they are predictive of the treatment but not the outcome. Variables have been centered and scaled.
  }
\end{figure}

Since the intuition furnished by the analytical results may not apply to our setting with a shrinkage prior, we further investigated whether these terms may be acting as instruments: We ran two OLS regressions, one for the treatment $Z$ and other for the outcome $Y$, on the controls in subset 1.2 (excluding the quadratic temporal trends for the state-level control variables, and including the treatment in the outcome regression). We then plotted the residuals against the control $\times$ year terms to look for evidence of an omitted partial effect.  These residual plots are shown in Figure~\ref{fig:reg-resid}.  There is some evidence of a quadratic trend in the treatment regression, while there is no noticeable trend in the residuals from the outcome regression.  While not a formal test, this is further evidence that these terms have the character of instrumental variables (given the other controls in the full model). The ability to identify potential instruments without refitting a model and without positing a model for $Z$ is a unique feature of our projected posterior approach to inference.



%% file: input/path.tex
\section{A path between control subsets}
\label{sec:path-betw-subm}


So far we have considered the case when there are specific subsets of controls that are of interest when estimating the treatment effect.  Alternatively we might be interested in learning which controls 
have the greatest influence on estimates of the treatment effect. Put simply, which terms can we omit from the model without significantly changing the estimated treatment effect? These control terms are behaving similar to purely prognostic variables, instruments, or noise. Removing these terms using the projected posterior can increase the precision and interpretability of the projected effect estimate. And while the behavior of the projected posterior alone cannot tell us which terms might correspond to prognostic, instrumental, or noise variables, it is suggestive. In this section we introduce a method to perform a guided exploration of alternative control specifications by adapting traditional regression tools.  In particular, we propose a backward stepwise approach which progressively excludes one control (or batch of controls) at a time. This produces a path of projected posteriors between control subsets, similar to our path in Section~\ref{sec:application} but ordered by the influence of the control over the projected effect estimate.  

In the simplest version of the algorithm we start from the full set of controls $\phi^{(0)} = (1,1,..,1)$ and at  step $d$ of the algorithm we consider a set of new candidate inclusion vectors $\Phi^{(q)}$ which is the set of all inclusion vectors obtained by dropping one control from $\phi^{(q-1)}$. For each $\phi'\in \Phi^{(q)}$ we compute the projected posterior $p(\tau_{\phi'}\mid Y)$ and compare it to the projected posterior under the full set of controls $\tilde\phi^{(0)}$ under some distance metric $d$. We then take
\begin{equation}
\phi^{(q)} = \argmin_{\phi'\in \Phi^{(q})} d(p(\tau_{\phi'}\mid Y), p(\tau^{(0)}\mid Y))\label{eq:stepdist}
\end{equation}
and iterate until the distance is too great, a desired level of sparsity is obtained, or all the controls are removed. 

As in backward stepwise regression, we can pre-specify a set of controls to always be included. We can also consider dropping {\em batches} of controls at each step, for example if we want to consider removing all the dummy variables corresponding to a categorical covariate in a single step. By default we take our distance metric to be the squared difference in posterior means:
\begin{align}%
  d(p(\tau_{\phi}\mid Y), p(\tau_{\phi'}\mid Y)) &:= (\tilde \tau_\phi -  \tilde \tau_{\phi'})^2. \label{eq:diff-mean}
\end{align}
This is a natural choice, as it emphasizes removing the controls that lead to the least ``bias'' (relative to the full set of controls). It is also easy to compute since at each step the posterior mean of $\tau_{\phi'}$ is available directly given the posterior mean of $\tau^{(0)}$ or $\tau^{(q-1)}$ using the formulas in Section~\ref{sec:proj-post-treatm} -- that is, we need not project each posterior sample for every candidate $\phi'$, only the posterior mean. However, other choices may make sense in particular contexts. For example, we might choose a distance that also considers the posterior variance to prioritize removal of terms that act as instruments rather than purely predictive variables, or we might choose an asymmetric function of the difference in posterior means that favors conservative projected estimates.



Finally, it is important to stress again that the goal here is not model or confounder selection in the usual sense. Recall that our inferences flow from the posterior under the full model, which represents the model uncertainty and includes all the confounders and controls. This procedure simply furnishes a set of controls with limited influence over the adjusted effect estimate in the projected posterior. In large samples this is essentially a function of the covariance of $Z$, $X$, and $Y$ (see Section~\ref{sec:properties}), so without additional information about the relationships between these variables it would be a mistake to use this path alone to determine causal relationships. However, we show in simulations below that the algorithm is generally successful at prioritizing removal of noise or instrumental variables early in the path when such variables exist. 

%% file: input/simulations.tex
\subsection{Simulation results of stepwise algorithm}
\label{sec:simulation-results}

Here we investigate the performance of our backward stepwise approach outlined in Algorithm 1.  We generate $n=1000$ observations from the model $Y = \tau Z + \beta_1 X_1 + \ldots + \beta_{14} X_{14} + \epsilon_i$, with $\tau = \beta_1 = \ldots \beta_{14} = 0.1$, and $\epsilon_i \sim \N(0, \sigma^2)$ with $\sigma^2 = 1$ known.  The exposure and covariates are sampled from the multivariate Gaussian $(Z, X_1, \ldots, X_7) \sim \N(0, \Sigma)$, with the covariance matrix defined element-wise by
\begin{align*}
  \Sigma_{kl} =  
  \begin{cases}
    1 & \text{if } k = l \\
    \rho^{k + l - 2} & \text{if } k \neq l
  \end{cases}
\end{align*}
for ${k, l}\in\{1, \ldots, 8 \}$ and $\rho = 0.7$.  Thus, of these covariates $X_1$ is most correlated with $Z$, and $X_7$ is least correlated with $Z$.  The remaining signal covariates $\{ X_8,\ldots, X_{14}\}$ are sampled independently from the standard Gaussian.  There are 11 additional noise variables (not involved in the true data generating mechanism) generated from a standard Gaussian, so in total there are $p = 25$ covariates.  The first seven covariates are confounders; the following seven covariates are prognostic variables (predictive of the outcome but not the treatment); the final eleven are noise variables.

We implement our posterior summarization technique as follows.  First, we obtain a posterior for $\tau$ and $\beta$ using the horseshoe prior \citep{horseshoe} with the likelihood parameterization from \cite{RIC} to prevent regularization-induced confounding.  Then, we use the backward stepwise algorithm above with the squared-difference in mean metric in Eq.~\eqref{eq:diff-mean} to iteratively remove covariates from the set of controls which appear to have the least confounding capacity, and report the corresponding path of projected posteriors for the treatment effect.

\begin{figure}[t!]
  \centering
  \includegraphics[width=1\textwidth]{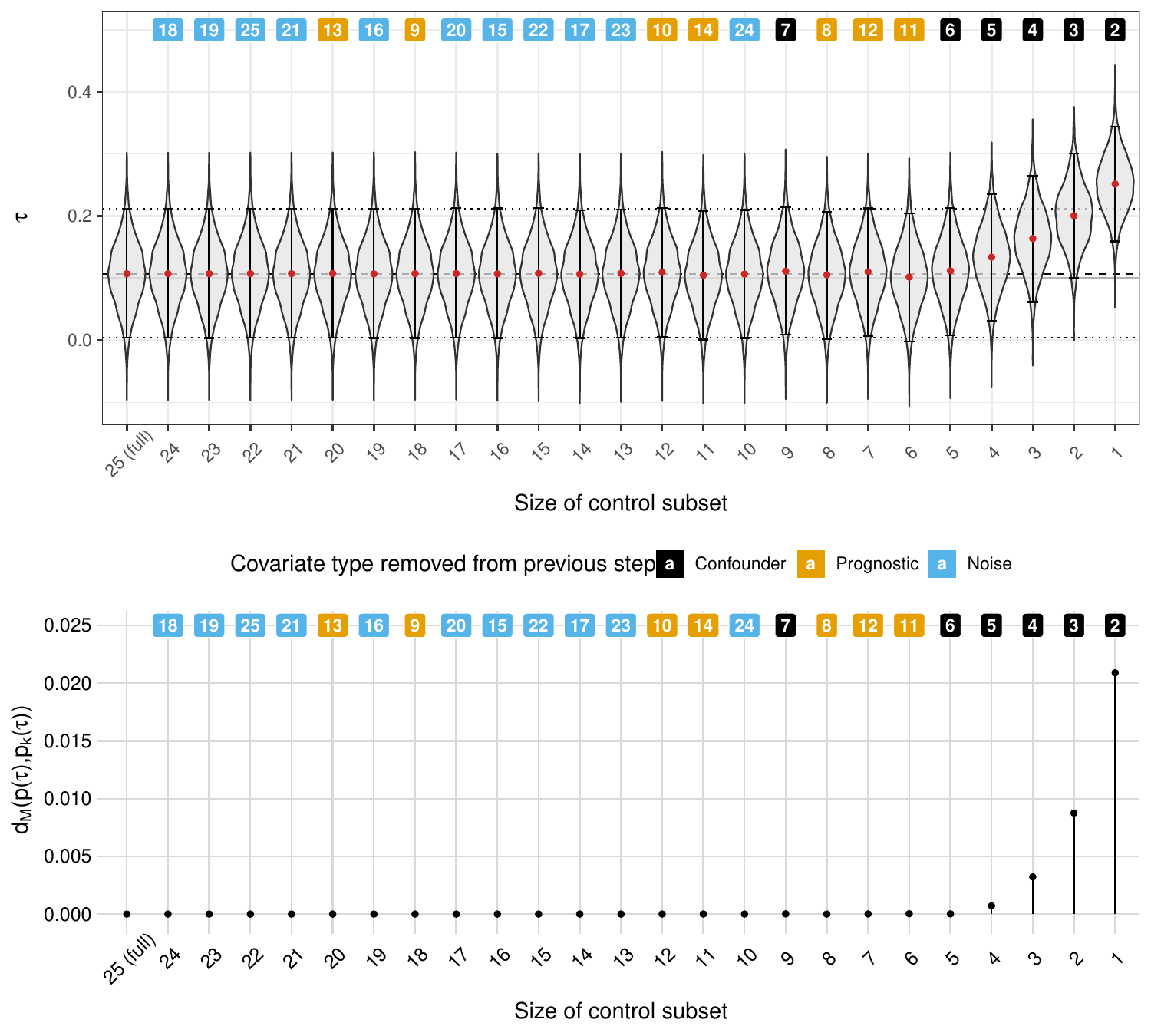}
  \caption{\label{fig:wang-stepwise-h}
    Stepwise summary path for simulated example using squared difference in mean for decision criterion. }
\end{figure}

Figure~\ref{fig:wang-stepwise-h} shows the results of the algorithm, displaying which covariates are removed at each step and the category of the covariate, i.e,. whether it is a confounder, prognostic, or noise variable.  We also show the projected posterior for each step, and the squared difference in posterior mean between the original posterior and the projected posterior, which is used as the decision criterion.  As we can see, we generally remove the noise variables first before removing the signal variables.  The confounding variables are removed only at the very end, with the exception of $X_7$ which is only very weakly correlated with the exposure, i.e.  $\text{cor}(Z, X_7) = 0.7^7 \approx 0.082$, and also correlated with $X_1, \ldots X_6$.  

\begin{figure}[t!]
  \centering
  \includegraphics[width=1\textwidth]{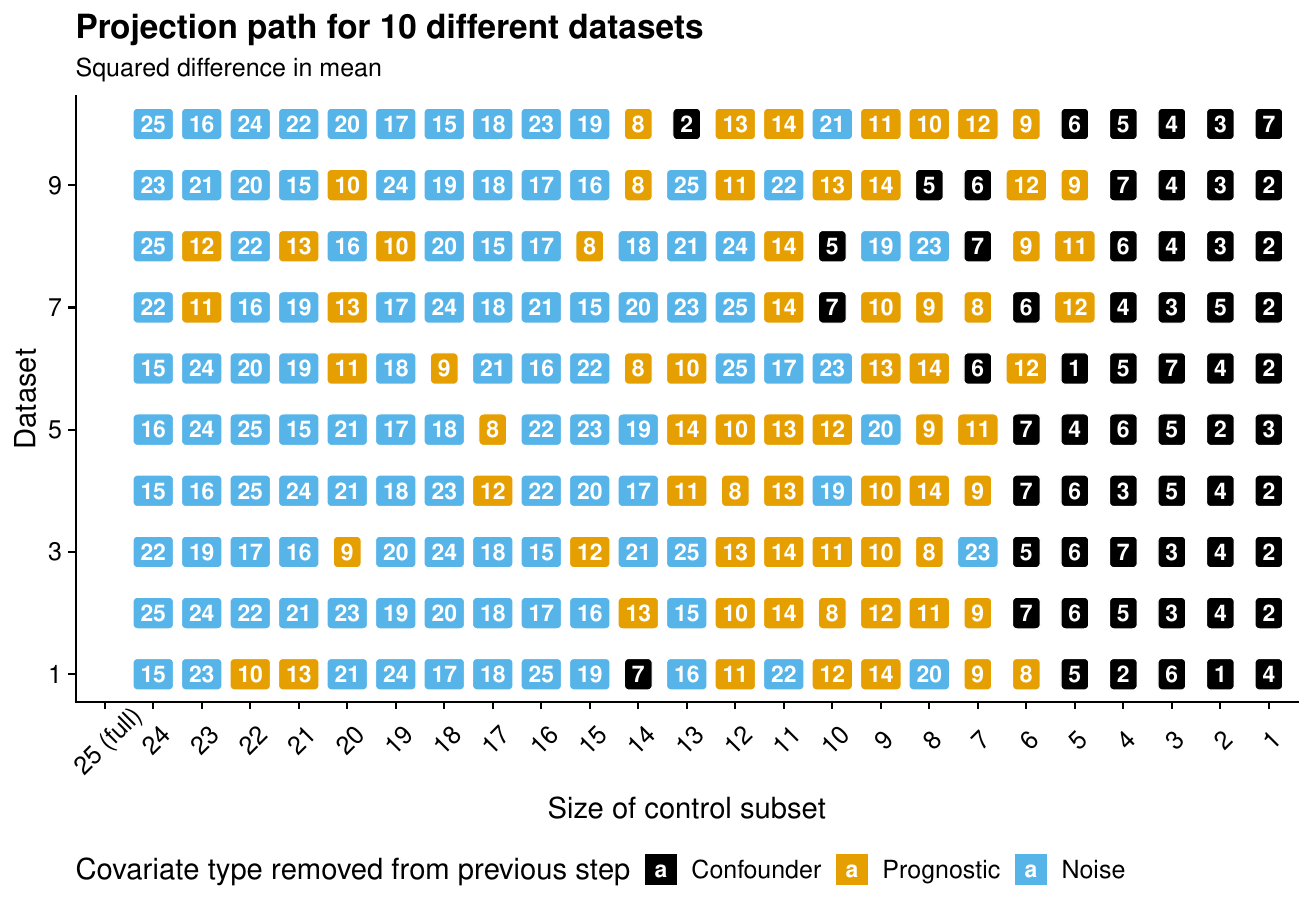}
  \caption{\label{fig:wang-ONECHAIN}
    Results of simulation example from Wang et al., showing summary path for one set of posterior samples from each of 10 different datasets generated from the same model. }
\end{figure}

As demonstrated in the plot of the squared difference in posterior mean, the projected posterior begins to significantly move away from the original posterior after covariate $X_5$ is removed.  This behavior is expected, as exclusion of confounding variables will bias the estimated treatment effect.  The last remaining covariate is $X_1$, the strongest confounder.  Figure~\ref{fig:wang-ONECHAIN} shows the control removal path for ten additional different datasets generated from the same model to show that this pattern is generally consistent across datasets.

\subsection{The impact of abortion on crime: A path between control subsets}

\begin{figure}[t!]
  \centering
  \includegraphics[width=1\textwidth]{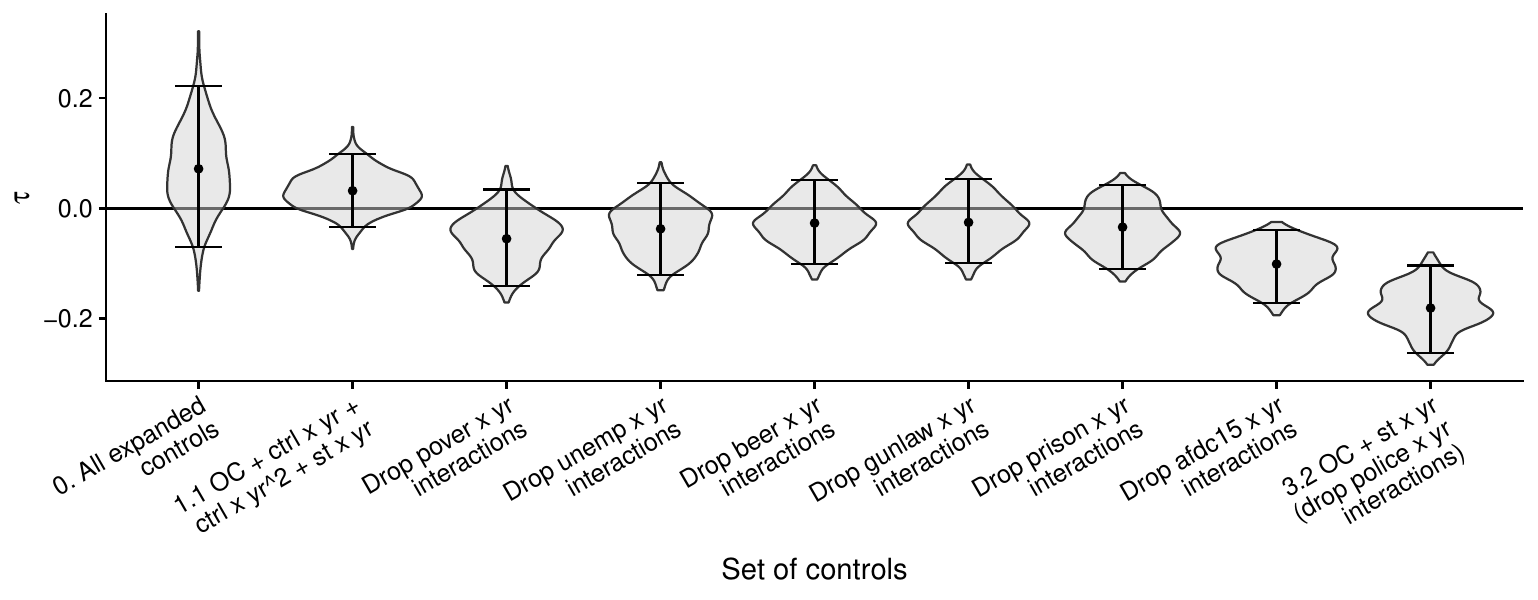}
  \caption{\label{fig:levitt-path} Application of stepwise algorithm for ranking controls in the Levitt data. }
\end{figure}

{
As noted in Section~\ref{sec:application}, the sign of the projected treatment effect estimate changes on introducing interactions between state dummies or the original controls and a quadratic temporal trend. That analysis raised the question of whether time-varying partial effects of the original controls were well-motivated. In this section we examine {\em which} of the original controls are most influential on the effect estimate when interacted with temporal trends.

Looking back to the top panel of Figure~\ref{fig:levitt-submodel-proj-comp}, we notice a marked change in the posterior between columns 1.2 and 3.2, which corresponds to the removal of interactions between state-level control variables and time (i.e. \texttt{prison} $\times$ \texttt{year} and \texttt{prison} $\times$ \texttt{year\^{}2}, etc.). On removing these time-varying effects we move from a projected posterior approximately centered on zero to a posterior centered near $-0.1$ with a 95\% credible interval excluding zero. 

To identify which of the original controls are most influential we use our stepwise algorithm, where at each step we consider removing a state-level control's interactions with the linear and quadratic time trends.
Figure~\ref{fig:levitt-path} shows the path between the controls in columns 1.2 and 3.2.  
The first pair of interaction terms dropped is state-level poverty, i.e. the terms for \texttt{pover} $\times$ \texttt{year} and \texttt{pover} $\times$ \texttt{year\^{}2}, though this projected posterior still overlaps 0.  After this, there is not much change in the projected posterior from dropping terms until the \texttt{afdc} terms are dropped (the sixth to be removed), which marks the first time the projected posterior credible interval does not encompass 0.  Finally, dropping the \texttt{police} temporal trends further lowers the projected posterior for the treatment effect.  

From this we can deduce that the \texttt{afdc} and \texttt{police} temporal trends are most important in the shift in the projected posteriors from control subsets 1.2 and 3.2.  This allows us to sharpen our discussion of the substantive justification for admitting time-varying partial effects of the controls, zeroing in on whether we should expect the partial effects of \texttt{afdc} and \texttt{police} to vary over time -- for example, were there changes over time in how police are {\em deployed} that made them more or less effetive in reducing crime on a per-capita basis (all else constant)? Were additional aid programs introduced, expanded, or cut over time in a way that modulates the effect of \texttt{afdc} over time? Answering these questions is beyond the scope of this paper, but our methods are a useful guide to which questions are worth asking in the first place. 
}

%% file: input/discussion.tex
\section{Discussion}
\label{sec:discussion}

In this paper we have introduced methods to conduct sensitivity analyses for treatment effect estimation in the common scenario where we assume a complete set of measured confounders but there is uncertainty about how they should enter the model.  This is an principled alternative to the heuristic method of conducting such a sensitivity analysis by fitting many models independently with the same outcome data.  In contrast to this frequently used ad-hoc approach, we use the data only once in computing the original posterior and then compute posterior summaries for the treatment effect when considering only restricted subsets of the controls.  

Using our methods we are able to shed new light on apparent sensitivity to control function specification in an important application. We find that effect estimates are in fact more robust than some authors originally suggested, that at least some of the expanded controls appear to be irrelevant, and that the difference in effect estimates can be eplained by a handful of additional controls, which supports a substantive debate about their merits from a modeling perspective.


There are many interesting avenues for future work: Existing applications of sensitivity analysis usually focus on  sensitivity to {\em unmeasured} confounding.  Here our focus instead has been on investigating sensitivity of treatment effect estimation with respect to control function specification.  Merging these two is an interesting area for future research.

We have considered the relatively simple setting of estimating homogenous effects with linear models. Extensions to estimating heterogeneous effects are straightforward in principle -- in this case we need only include interactions between $Z$ and some of the controls, in which case $\tau$ is extended to a vector of coefficients parameterizing the heterogeneous effects. All the results in Section~\ref{sec:proj-post-treatm} are easily adapted to a vector $\tau$. Extensions to nonlinear response models (e.g. generalized linear models) are possible as well, but there are new considerations when choosing the distance metric in the posterior summary which defines the projection operator. 

Finally, one might be interested in summarizing models with fully nonparametric priors over the control function (or heterogeneous effects) rather than a hand-selected set of nonlinear and interaction terms. Conceptually nothing changes -- one can motivate the projected posterior in the same way, and compute it by regressing posterior samples of $\hat{y}$ on $\Wt$. However in this case it is less clear what it means for a control specification to be ``nested'' under the full model, and non-nested summaries introduce new complications.  It is nontrivial to even define a posterior summary that includes only a subset of the confounding variables but is as ``close'' as possible to the original model when a nonparametric tree prior like Bayesian causal forests \citep{bcf} is used as the full model. These are interesting and important areas for future work.